\documentclass[12pt]{article}
\usepackage{geometry}
\usepackage[T1]{fontenc} 
\usepackage{textcomp}
\usepackage{times}
\usepackage[scaled=0.92]{helvet}

\usepackage[auth-lg,affil-sl]{authblk}
\usepackage{amsmath}
\usepackage{amssymb}
\usepackage{graphicx}
\usepackage{subfigure}

\title{\textbf{A Compact Codimension Two Braneworld}\\ \textbf{with Precisely One Brane}}

\author{Nikolas Akerblom\footnote{{\tt nikolasa@nikhef.nl}}}
\affil{Nikhef Theory Group\\Amsterdam, The Netherlands}

\author{Gunther Cornelissen\footnote{{\tt g.cornelissen@uu.nl}}}
\affil{Department of Mathematics\\Utrecht University, The Netherlands}

\begin{document}

\maketitle

{\abstract{{ \noindent Building on earlier work on football shaped extra dimensions, we construct a compact codimension two brane\-world with precisely one brane. The two extra dimensions topologically represent a $2$-torus which is  stabilized by a bulk cosmological constant and magnetic flux. The torus has positive constant curvature almost everywhere, except for a single conical singularity at the location of the brane. In contradistinction to the football shaped case, there is no fine-tuning required for the brane tension. We also present some plausibility arguments why the model should not suffer from serious stability issues.\\

\begin{flushright} Nikhef 2010-009 \end{flushright}}}}

\newpage

\section{Introduction}
Ever since the Arkani-Hamed--Dimopoulos--Dvali (ADD) proposal \cite{ArkaniHamed:1998rs} (see \cite{Antoniadis:1998ig} for the first stringy realization) to explain the relative weakness of gravity, a great deal of effort has been expended on the study of theories with large extra dimensions. One particular approach is that of codimension two braneworlds \cite{Carroll:2003db,Navarro:2003vw}. In these models the observed four-dimensional universe is pictured to be stuck on a stack of $3$-branes embedded in a six-dimensional spacetime with bulk cosmological constant and magnetic flux in the two extra dimensions. Insofar as the extra dimensions are warped, there is a certain similarity to the RS models \cite{Randall:1999ee,Randall:1999vf}.

To have flat branes, the magnetic flux has to be fine-tuned so as to cancel the \emph{brane} warping induced by the bulk cosmological constant, and the extra dimensions topologically represent a sphere which is stabilized by the magnetic flux. As already pointed out in \cite{Carroll:2003db}, in addition to the fine-tuning between the cosmological constant and the magnetic flux, there is at least one other source of fine-tuning in this scenario: For mathematical reasons, it is necessary to have at least two branes whose tensions then necessarily are \emph{exactly} equal. The resulting geometry in the extra dimensions generically\footnote{Compare Section \ref{football}.} is that of an American-style football.

In the literature there already have been attempts to remove the necessary fine-tuning of the brane tensions by replacing the extra-dimensional sphere with a non-compact tear\-drop shaped space \cite{Kehagias:2004fb}. In this background geometry it is possible to have precisely one brane and therefore no issue of fine-tuning arises for its tension.

Generally speaking, `compactification' on a non-compact space\footnote{There is an extensive literature on `non-compact compactifications,' a small sample of which is \cite{Rubakov:1983bz,Nicolai:1984jga,RandjbarDaemi:1985wg,GellMann:1984mu,GellMann:1985if,Kehagias:2000dga,Cohen:1999ia}.} raises various stability questions. Even though these questions may be settled for the teardrop, we felt it worthwhile to construct a \emph{compact} model with precisely one brane. The purpose of this letter is to outline our construction.\footnote{Another approach was taken in \cite{Lee:2004vn}, where the sigma-model leading to the teardrop was replaced so as to obtain a compact space supporting precisely one brane.} 

\subsection{Sketch of the model}
A brief description of our model is as follows. Spacetime is assumed to be six-dimensional and there is a bulk cosmological constant $\Lambda>0$. The two extra dimensions are threaded by magnetic flux and form what we shall call \emph{Olesen space}.

Topologically speaking, Olesen space is just the two-torus $T^2$. In particular, it is compact. The bulk cosmological constant forces the space to have constant positive curvature almost everywhere, \emph{except} at a single conical singularity. In accordance with the Gauss-Bonnet formula, we may ascribe a negative infinity of curvature to this conical singularity.

We now think of the universe as a single space-filling $3$-brane in this background. The brane is point-like in the extra dimensions and sits at the conical singularity.\footnote{Physically speaking, this description is of course slightly backwards, as the brane itself is the reason for the conical singularity.} By necessity, the brane must have negative tension but this tension can vary within a certain interval, so there is no fine-tuning here.

\paragraph{Organization.}
In order to flesh out the description just given, in the next section we first review the required general codimension two braneworld construction. In Section \ref{solutions} we then remark on certain sometimes overlooked geometric features of the football shaped extra dimensions before proceeding to Olesen space. Finally, in Section \ref{discussion} we summarize some salient points of our construction and remark on directions for future work.

\section{Review of general construction}\label{review}
We review here the general construction for codimension two braneworlds \`{a} la \cite{Carroll:2003db,Navarro:2003vw} (cf.\ also \cite{Redi:2004tm}).
In these models, spacetime has the form
\begin{equation}
\text{4D Minkowski space}\times M^2,
\end{equation}
where $M^2$ is some compact space of finite volume with coordinates $u, v$.

The metric is
\begin{equation}
ds^2=\eta_{\mu\nu}\,dx^\mu dx^\nu+e^{2\phi(u,v)}(du^2+dv^2),
\end{equation}
or in matrix notation
\begin{equation}
g_{MN}=
\begin{bmatrix}
\eta_{\mu\nu} & 0\\
0 & e^{2\phi}\delta_{uv}
\end{bmatrix},
\end{equation}
where $M,N=0,1,2,3,u,v$.

In writing the part of the metric for $M^2$,
\begin{equation}\label{confcoords}
e^{2\phi(u,v)}(du^2+dv^2),
\end{equation}
we have made use of the fact that $2$-space is conformally flat.

As to the `physical' ingredients, we have a (bulk) cosmological constant
\begin{equation}
\Lambda>0,
\end{equation}
and $M^2$ is threaded by a constant magnetic flux with field strength\footnote{It is easy to see that this is a solution of the vacuum Maxwell equations. Note that for purely cosmetic reasons we have rescaled $B_0$ by a factor of $\sqrt{2}$ relative to \cite{Carroll:2003db}.}
\begin{equation}
F=\sqrt{2}\,B_0\,e^{2\phi}\, du\wedge dv.
\end{equation}
In addition, there is a stack of space-filling $3$-branes, see Table \ref{branes}.
\begin{table*}[t]
\caption{Branes. $\times$ indicates a `filled' dimension, $\bullet$ indicates that the branes are `pointlike' there.}
\begin{center}
\begin{tabular}{c|c|c|c|c|c|c}
$M, N=$ & $0$ & $1$ & $2$ & $3$ & $u$ & $v$\\
\hline
& $\times$ & $\times$ & $\times$ & $\times$ & $\bullet$ & $\bullet$
\end{tabular}
\end{center}
\label{branes}
\end{table*}

At the level of detail we are working here, that is, just getting the basic construction right and not worrying about phenomenology on the branes, the only equation to satisfy is the Einstein equation (we put $M_6^4=1$)
\begin{equation}\label{einstein}
R_{MN}-\frac{1}{2}Rg_{MN}=T_{MN}.
\end{equation}

The energy momentum tensor consists of three parts,
\begin{equation}
T_{MN}=T_{MN}^\Lambda+T_{MN}^F+T_{MN}^\text{branes},
\end{equation}
where
\begin{align}
T_{MN}^\Lambda&=-\Lambda
\begin{bmatrix}
\eta_{\mu\nu} & 0\\
0 & e^{2\phi}\delta_{uv}
\end{bmatrix},\\
T_{MN}^F&=-B_0^2
\begin{bmatrix}
\eta_{\mu\nu} & 0\\
0 & -e^{2\phi}\delta_{uv}
\end{bmatrix},\\
T_{MN}^\text{branes}&=-e^{-2\phi}
\begin{bmatrix}
\eta_{\mu\nu} & 0\\
0 & 0
\end{bmatrix}\sum^\mathcal{N}_{n=1} \sigma_n\, \delta^2(u_n,v_n).
\end{align}
The delta-functions in the brane energy momentum tensor correspond to the location of the various branes in the extra-dimensional space $M^2$ and $\sigma_n$ is the tension of brane $n$.

Working out the components of the Einstein equation, we find:
\begin{align}
e^{-2\phi} \,\eta_{\mu\nu}\,\Delta\phi
&=(-\Lambda-B_0^2)\,\eta_{\mu\nu}
-e^{-2\phi}\,\eta_{\mu\nu} \sum^\mathcal{N}_{n=1} \sigma_n\, \delta^2(u_n,v_n)\label{firsteinstein},\\
0 &=-\Lambda e^{2\phi}\delta_{uv}+B_0^2 e^{2\phi} \delta_{uv},\label{secondeinstein}
\end{align}
where
\begin{equation}
\Delta=\partial_u^2+\partial_v^2
\end{equation}
is the Laplacian.

Equation \eqref{secondeinstein} is equivalent to
\begin{equation}
\Lambda=B_0^2.
\end{equation}
This equation is an expression of the fine-tuning between the bulk cosmological constant and the magnetic flux generically present in codimension two braneworlds. We assume it to be satisfied. Thus, substituting $\Lambda$ for $B_0^2$ in equation \eqref{firsteinstein} and rearranging, we arrive at
\begin{equation}\label{liouville}
\Delta\phi=-2\Lambda\, e^{2\phi}-\sum^\mathcal{N}_{n=1} \sigma_n\, \delta^2(u_n,v_n),
\end{equation}
which we recognize as the famous Liouville equation. With $z=u+iv$, its general solution for $\Lambda>0$ is given by
\begin{equation}\label{gensol}
e^{2\phi}=\frac{2}{\Lambda}\frac{|f'(z)|^2}{(1+|f(z)|^2)^2},
\end{equation}
with $f(z)$ some (in general multivalued) complex function. Now, typically the data in the Liouville equation---total number of branes $\mathcal{N}$, location of the branes $(u_n,v_n)$, tensions $\sigma_n$---are not entirely free, but depending on the topology of $M^2$ are constrained. For instance, for $M^2=S^2$ the total number of branes $\mathcal{N}$ necessarily is $\geq2$. We will come back to this below.

Due to the well-known geometric meaning of the Liouville equation, we immediately read off from \eqref{liouville} that $M^2$, outside of the delta-function singularities, has constant curvature
\begin{equation}
K=2\Lambda.
\end{equation}

The Gauss-Bonnet formula for a compact surface $M^2$ with conical singularities with `weights' $\beta_n>-1$  and Euler characteristic $\chi(M^2)$ reads \cite{Troyanov2}
\begin{equation}
\frac{1}{2\pi}\int_{M^2}K\,dA=\chi(M^2)+\sum\beta_n.
\end{equation}
Applied to our braneworld case, we find
\begin{equation}\label{gaussb}
\frac{\Lambda}{\pi}\int e^{2\phi}\,dudv=\chi(M^2)-\frac{1}{2\pi}\sum\sigma_n,
\end{equation}
whence it follows that a brane of tension $\sigma_n$ locally gives rise to some kind of cone geometry with angle $\theta_n$, such that
\begin{equation}
\sigma_n=2\pi-\theta_n,
\end{equation} 
that is, $\sigma_n$ represents the angular defect (for $\sigma_n>0$) or excess (for $\sigma_n<0$) of the local cone geometry. We also remark here that the volume of $M^2$ is given by
\begin{equation}\label{volume}
\int e^{2\phi}\,dudv,
\end{equation}
and from \eqref{gaussb} we see that this is entirely determined by $\Lambda$, $\chi(M^2)$, and $\sum\sigma_n$. For fixed cosmological constant $\Lambda$ and Euler characteristic $\chi(M^2)$, the question whether a model can decompactify and hence is stable or unstable, as it were, is entirely determined by the brane tensions $\sigma_n$.

From \eqref{gaussb} it follows that necessarily\footnote{We restrict ourselves to orientable $M^2$. }
\begin{equation}\label{tens}
\sum\sigma_n<0\quad\text{for } M^2\neq S^2,
\end{equation}
as then the Euler characteristic $\chi(M^2)\leq0$ and the left hand side of \eqref{gaussb} is positive.

The upshot of this section can be briefly summarized thus: To construct a codimension two braneworld, all we need to do is solve the Liouville equation \eqref{liouville}. The general solution is given by \eqref{gensol} but one still has to do a little extra work to select an acceptable $f(z)$ in order to satisfy the Liouville equation with given (compatible) data. Nevertheless, very many solutions can be written down explicitly, for instance
\begin{equation}
M^2=S^2:\quad f(z)=\text{a rational function}
\end{equation}
and
\begin{equation}
M^2=T^2:\quad f(z)=\text{an `}\Omega\text{-quasi-elliptic function'}
\end{equation}
give for our purposes acceptable solutions of the Liouville equation \cite{Horvathy:1998pe,Akerblom:2009ev}. Let us hasten to add that these are not the most general solutions in the \emph{present} context, as the $f(z)$ in both cases are single-valued in the plane, but for example, $f(z)=z^k$ $(0<k<1)$ also leads to a viable solution.\footnote{In references \cite{Horvathy:1998pe,Akerblom:2009ev}, the Liouville equation was studied in connection with the Jackiw-Pi model. There, because of its relation to the gauge field, $f$ had to be single-valued in the plane but that is not the case in the current theory.} Also, most of these solutions are appropriate only for a large number of branes in the construction.

On the sphere $S^2$ it is known\footnote{To the physicist this is quite obvious from the following argument. Suppose we look for a single-brane solution of the Liouville equation on the sphere. We expect this solution to be radially symmetric and thus make the ansatz $f(z)=z^k$ in equation \eqref{gensol}. It is then easy to check that the resulting metric has a singularity at $z=0$ \emph{and} $z=\infty$.} \cite{Troyanov2} that the minimal number of branes allowed is \emph{two} and this is the model proposed in \cite{Carroll:2003db,Navarro:2003vw}. A long time ago, Olesen \cite{Olesen:1991df,Olesen:1991dg} found a solution of the Liouville equation on the square torus with precisely \emph{one} singularity,\footnote{In \cite{Akerblom:2009ev} this was generalized to tori of arbitrary shape.} corresponding to precisely one brane in our braneworld. This solution is a bit subtle, since here the function $f$ itself is not single-valued on $T^2$ (it is not an elliptic function with respect to $T^2$).

We now turn to the discussion of these minimal cases.

\section{Minimal solutions}\label{solutions}
\subsection{The football and related geometries}\label{football}
In the case where the extra-dimensional space $M^2$ is homeomorphic to the sphere $S^2$, all axisymmetric  solutions with two branes were given in \cite{Carroll:2003db}. In the notation set up in the previous section, these solutions are generated by\footnote{Strictly speaking, only the solutions with $0<k<1$ were discussed in \cite{Carroll:2003db}. They correspond to positive tension branes.}
\begin{equation}
f(z)=z^k\quad (k>0, k\neq1),
\end{equation}
viz.
\begin{equation}
e^{2\phi}=\frac{2}{\Lambda}\,\frac{k^2\,|z|^{2k-2}}{(1+|z|^{2k})^2}.
\end{equation}
We exclude the case $k=1$, which corresponds to the usual round metric on the sphere without point sources (indeed, $\sigma_1=\sigma_2=0$ for $k=1$, see below).

It is not difficult to check that with this geometry, there is one brane sitting at $z=0$ and one brane sitting at $z=\infty$, and that both branes have the same tension,
\begin{equation}
\sigma_1=\sigma_2=\sigma=2\pi(1-k).
\end{equation}
In fact, it is a rephrasing of a theorem of differential geometry \cite{Troyanov} to say that for all solutions on the sphere $S^2$ with precisely two branes, the branes must have exactly the same tension.

One surprising fact, which is not so often stressed, is that even though generically the branes are antipodal on $S^2$, whenever
\begin{equation}
\sigma_1=\sigma_2=-2\pi,-3\pi,-4\pi,\dots,
\end{equation}
there also exist solutions where the branes are \emph{not} antipodal \cite{Troyanov}.
Indeed, for tension $\sigma_1=\sigma_2=2\pi(1-k)$ with $k$ an integer $>1$, these solutions arise as pullbacks of the standard metric on the sphere by a ramified covering $f\colon S^2\rightarrow S^2$ of degree $k$, totally ramified over exactly two points. If we view $S^2$ as the Riemann sphere $\hat{\mathbf{C}}$ and choose a complex coordinate $ f\colon \hat{\mathbf{C}}_w \rightarrow \hat{\mathbf{C}}_z$ such that $z=\infty$ corresponds to one of the ramification points, then the map $f$ is described algebraically by an equation of the form $w^k=z-p,$ where $p$ is the $z$-coordinate of the second ramification point.

One can summarize the situation as follows: if one has two (and only two) branes of varying (equal) tension, they are forced to remain at a fixed distance $\pi/\sqrt{2 \Lambda}$ on the sphere, but for certain `instantaneous' values of the tension, the branes can start to wander around freely on the sphere. We do not know of a simple physical reason for this phenomenon.

\subsection{Olesen space}
We have seen that in the case where $M^2$ is homeomorphic to the sphere $S^2$, the minimal number of branes is two and that then the brane tensions have to be infinitely fine-tuned to be exactly equal. We are not going to debate whether this is natural or not but we nevertheless think it is interesting to point out that in the case where $M^2$ is homeomorphic to the torus $T^2$, we can write down an explicit model with precisely one brane. It follows from \eqref{tens} that the price we have to pay is the admission of negative tension branes into the theory. This is not terribly problematic, as negative tension branes are a crucial ingredient of the Randall-Sundrum models \cite{Randall:1999ee,Randall:1999vf} and are also quite common in string theories (O-planes). In addition, in \cite{Parameswaran:2007cb} it was found that stability of axisymmetric 6D SUGRA braneworlds with $M^2$ homeomorphic to $S^2$ is actually easier to obtain once one admits negative tension branes.

Two tori in conformal coordinates \eqref{confcoords} can still differ in shape (be skinny or fat) or what amounts to the same, differ in their complex structure $\tau$.

Thus, we may assume that
\begin{equation}
T^2=\mathbf{C}/(\mathbf{Z}+\tau\mathbf{Z}),
\end{equation}
where $\tau$ is a non-real complex number (which can be chosen in the fundamental region of the modular group).

The coordinates $(u,v)$ in \eqref{confcoords} now range over the parallelogram $F\subset\mathbf{R}^2\equiv\mathbf{C}$ spanned by $1$ and $\tau$, that is
\begin{equation}
(u,v)\in F=\{s+t\tau\,|\,0\leq s,t \leq 1\},
\end{equation}
where parallel edges are identified in the usual way.

We want to write down an explicit solution to the Liouville equation \eqref{liouville} corresponding to a single brane on the torus and the formulas are least messy when we assume $\tau=i$. Thus, for now, assume
\begin{equation}
\tau=i.
\end{equation}

Put $z=u+iv$,
\begin{equation}
\wp(z)\equiv\wp_{2,2i}(z),
\end{equation}
(the Weierstrass p-function associated with the \emph{doubled} lattice $2\mathbf{Z}+2i\mathbf{Z}$), and $e_1=\wp(1)$.

Then
\begin{equation}\label{ole}
\exp\big({2\phi_\mathcal{O}(u,v)}\big)=\frac{2}{\Lambda}\,\frac{|\wp'|^2\,|e_1|^2}{(|e_1|^2+|\wp|^2)^2}
\end{equation}
is a solution to the Liouville equation \eqref{liouville} on $T^2$ with $\tau=i$.\footnote{Note that \eqref{ole} is of the form \eqref{gensol} with $f(z)=\wp(z)/e_1$.}

As this solution was obtained by Olesen \cite{Olesen:1991df,Olesen:1991dg} (albeit in a different context), we shall refer to $T^2$ equipped with the metric
\begin{equation}
\exp\big({2\phi_\mathcal{O}(u,v)}\big)\,(du^2+dv^2)
\end{equation}
as \emph{Olesen space}. For the general $\tau\neq i $ case, one can adapt the results of \cite{Akerblom:2009ev} but we choose not to do so here. Note however that unless we state the contrary, all statements below hold for the $\tau\neq i $ case as well.

Obviously, Olesen space has positive constant curvature $K=2\Lambda$ almost everywhere, and it is not hard to verify that it has a single conical singularity at $(u,v)=(0,0)$, corresponding to a brane  of tension
\begin{equation}
\sigma=-2\pi.
\end{equation}

\begin{figure*}[t]
\centering
\subfigure[]{
\includegraphics[width=5.6cm]{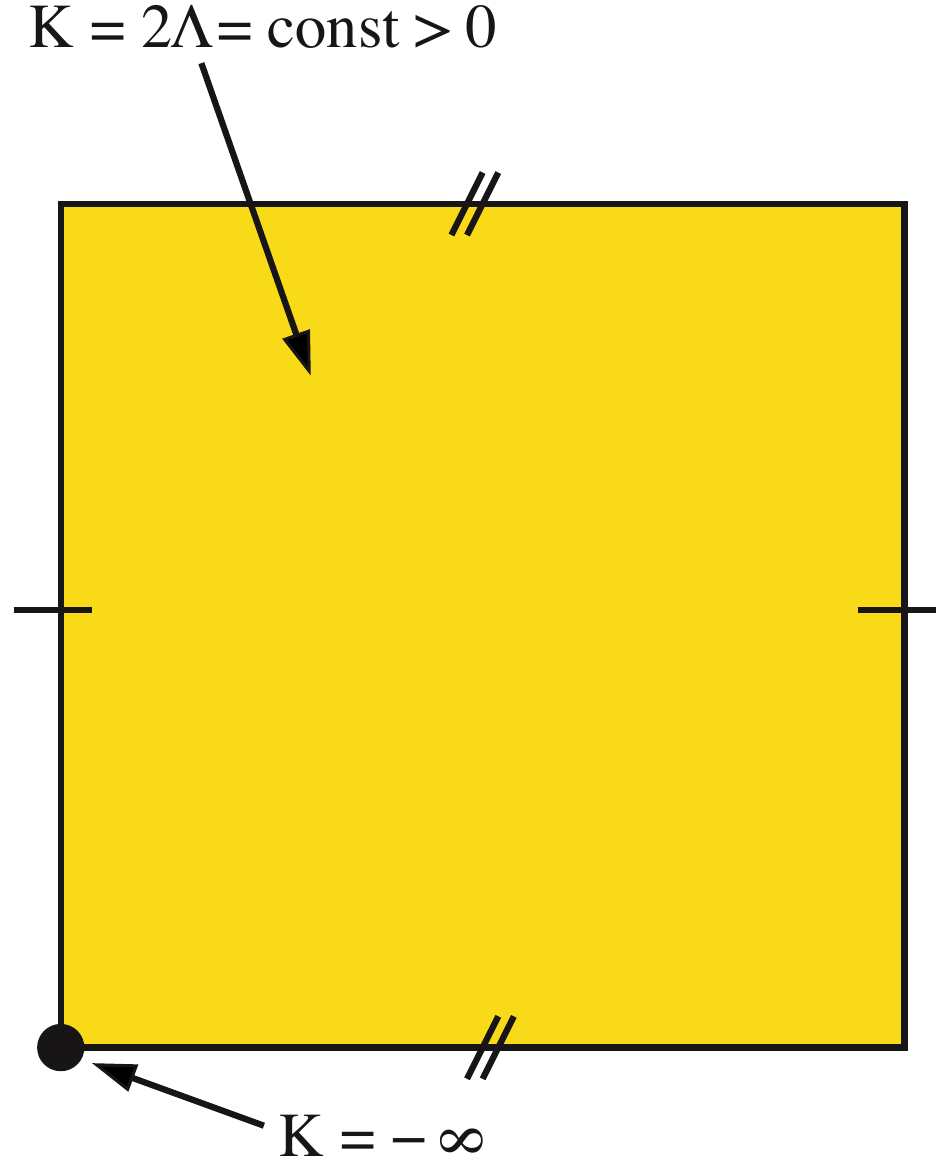}
\label{fig:ident}
}
\subfigure[]{
\includegraphics[width=7.3cm]{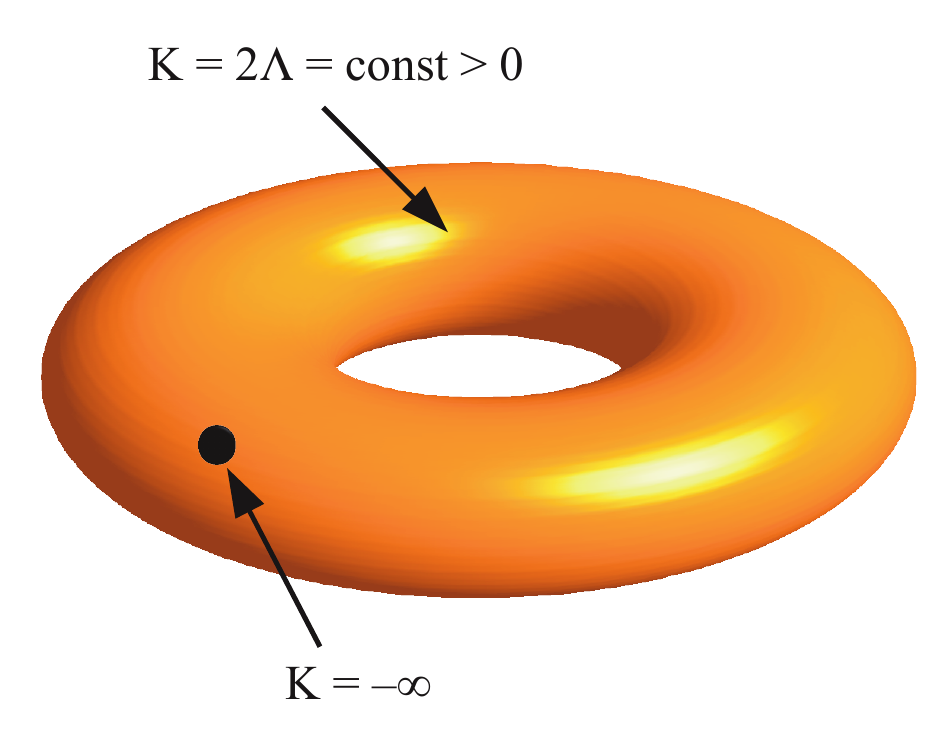}
\label{fig:torus}
}
\caption{Olesen space. The underlying topological space is just the $2$-torus, which in \subref{fig:ident} is obtained by identification on the sides of a square. \subref{fig:torus} shows a (non-isometric!) embedding in $3$-space. The curvature $K$ is everywhere constant positive and equal to $2\Lambda$, \emph{except} at a single conical singularity $\bullet$ to which, in accordance with the Gauss-Bonnet formula, we may ascribe the curvature $-\infty$. This conical singularity is also the position of our single $3$-brane in the extra dimensions.}\label{olesenspace}
\end{figure*}

Figure \ref{olesenspace} attempts a visual representation of Olesen space.

From results in differential geometry \cite{Troyanov2}, it follows that there exists a continuum of solutions with a single brane of tension $\sigma$ with\footnote{The `Trudinger constant' of \cite{Troyanov2} equals $2$ in the case at hand.}
\begin{equation}\label{interval}
-4\pi<\sigma<0.
\end{equation}

We do not know whether these can be expressed in `closed form,' even though it should not be too difficult to obtain approximate solutions by perturbing around the Olesen solution. In any case, it is very difficult to obtain results for the `critical' ($\sigma=-4\pi$) or the `supercritical' ($\sigma<-4\pi$) case.

We add a remark about the algebraic geometry of such solutions. It is known that a conformal metric of constant curvature outside a divisor gives rise to a so-called \emph{projective connection}. In the case of a torus with one brane of tension $\sigma$, any two such projective connections differ by a quadratic differential with at most a pole at the singular point $p$ of the metric \cite{Troyanov}, viz., an element of
\begin{equation}
H^0(T^2, \Omega_{T^2}^{\otimes 2}(p))=H^0(T^2,\mathcal{O}_{T^2}(p))=\mathbf{C},
\end{equation}
since the only functions on a torus with at most one pole are constant. In our case, this says that \emph{if} there is a solution to the Liouville equation, then its associated projective connection is unique up to scaling.

\paragraph{General $\tau$ and stability.}
Recent advances in the mathematical literature (cf.\ \cite{Taiwan} and references therein\footnote{Also compare \cite{Berman}.}) indicate that the solvability of the Liouville equation depends in a very subtle way on the complex modulus $\tau$ of the torus. Most results concern the `mean field equation,' which is the case where there is a further relation between the cosmological constant and the brane tension, namely
\begin{equation}
\Lambda=-\sigma/2.
\end{equation}
In this case, it is known from degree theory 
that a solution always exists if $\sigma$ is \emph{not} of the form  $\sigma=-4 \pi k$ for a positive integer $k \geq 1$. In the first exceptional case, $\sigma=-4\pi$, it is known that there is at most one solution. However, the amazing result is that the \emph{existence} of such a solution for $\sigma=-4\pi$ depends on the complex parameter $\tau$, more precisely, it exists only if the Green's function of the torus has more than three critical points. For example, it is known that there is no solution for $\sigma=-4\pi$ on the square torus $\tau=i$, but there is a solution on the rhombic torus $\tau = e^{\pi i / 3}$.

We can now use these results to argue for the stability of our model in a particular instance. Suppose we fix $\tau=i$ and $\Lambda=2\pi$. Further, assume that we start with a brane of tension, say, $\sigma=-2\pi$. The torus thus initially has volume $-\sigma/2\Lambda=1/2$ (cf.\ the discussion around \eqref{volume}). The decompactification limit of our model corresponds to $\sigma\to-\infty$. But because of the `gap' at $\sigma=-4\pi$, this limit cannot be taken in a smooth way! In other words, the model appears stable against decompactification in the described circumstances.

It would be a very interesting problem to study the existence of gaps like the one just described throughout the (super)critical region for general $\Lambda$ and $\sigma$, both for fixed complex structure $\tau$, and for varying $\tau$.

\section{Discussion}\label{discussion}
In the context of \cite{Carroll:2003db,Navarro:2003vw} we have constructed a compact codimension two braneworld with precisely one brane. Spacetime is six-dimensional with two dimensions compact. The compact space, dubbed Olesen space, topologically is the $2$-torus, threaded by magnetic flux. Olesen space has constant curvature $2\Lambda$ (where $\Lambda>0$ is the bulk cosmological constant) except for a single conical singularity, representing the position of the brane in the extra dimensions.

In the case where the extra dimensions topologically represent a sphere, the minimum number of branes is two and their tensions have to be infinitely fine-tuned to be exactly equal. This possibly unwanted feature is absent in our model. In fact, the brane tension can assume any value within a certain interval.

We briefly discuss the stability of our model. We found that this is intimately linked to the existence of gaps in the so-called (super)critical region $\sigma\leq-4\pi$ of the Liouville equation. Using recent results in mathematics, we argued for the absence of a smooth decompactification limit in a particular instance.

In the general case, we do not know whether or not solutions to the Liouville equation exist for all negative values of the tension, where in the case of existence, one possibly would have to deal with more subtle stability problems.

Quite apart from these basic issues, there is the task of building a viable phenomenology on top of the geometric background given here. For instance, in \cite{Schwindt:2005fm} the authors reach the conclusion that \emph{``a six-dimensional universe with two branes in the ``football-shaped'' geometry leads to an almost realistic cosmology.''} It would certainly be interesting to know whether this is also true for the Olesen space model.\footnote{Note that the changed topology of the internal space clearly has some import for, say, the spectrum of Kaluza-Klein states.}

\paragraph{Acknowledgments.}
It is our pleasure to thank F. Beukers, D. George, P. Horvathy, E. Plauschinn, M. Postma, B. Schellekens, and J.-W. van Holten for stimulating discussions, and in particular D. L\"ust for valuable comments on the manuscript. The work of NA has been supported by the Dutch Foundation for Fundamental Research on Matter (FOM).

\bibliographystyle{utphys}	
\bibliography{paper}	

\end{document}